\pdfoutput=1
\documentclass[aps,reprint,superscriptaddress]{revtex4-2}
\usepackage{amsmath,amssymb,amsfonts}
\usepackage{newtxtext}
\usepackage[varvw]{newtxmath}
\usepackage{graphicx}
\usepackage[dvipsnames]{xcolor}
\usepackage[colorlinks=true,citecolor=blue]{hyperref}
\usepackage{bm}

\graphicspath{{figs}}

\begin{document}

\title{Ashkin-Teller phase transition and multicritical behavior in a classical monomer-dimer model}

\author{Satoshi Morita}
\thanks{Current address: Faculty of Science and Technology, Keio University, Yokohama, Japan}
\email[]{smorita@keio.jp}
\affiliation{
  Institute for Solid State Physics,
  The University of Tokyo, Kashiwa, Chiba 277-8581, Japan
}

\author{Hyun-Yong Lee}
\affiliation{
  Department of Applied Physics, Graduate School,
  Korea University, Sejong 30019, Korea
}
\affiliation{
  Division of Display and Semiconductor Physics,
  Korea University, Sejong 30019, Korea
}
\affiliation{
  Interdisciplinary Program in E·ICT-Culture-Sports Convergence,
  Korea University, Sejong 30019, Korea  
}

\author{Kedar Damle}
\affiliation{
  Department of Theoretical Physics,
  Tata Institute of Fundamental Research, Mumbai 400 005, India
}

\author{Naoki Kawashima}
\affiliation{
  Institute for Solid State Physics,
  The University of Tokyo, Kashiwa, Chiba 277-8581, Japan
}
\affiliation{
  Trans-scale Quantum Science Institute,
  The University of Tokyo, Bunkyo-ku, Tokyo 113-0033, Japan
}

\date{\today}

\begin{abstract}
We use Monte Carlo simulations and tensor network methods to study a classical monomer-dimer model on the square lattice with a hole (monomer) fugacity $z$, an aligning dimer-dimer interaction $u$ that favors columnar order, and an attractive dimer-dimer interaction $v$ between two adjacent dimers that lie on the same principal axis of the lattice. The Monte Carlo simulations of finite size systems rely on our grand-canonical generalization of the dimer worm algorithm, while the tensor network computations are based on a uniform matrix product ansatz for the eigenvector of the row-to-row transfer matrix that work directly in the thermodynamic limit. The phase diagram has nematic, columnar order and fluid phases, and a nonzero temperature multicritical point at which all three meet. For any fixed $v/u < \infty$, we argue that this multicritical point continues to be located at a nonzero hole fugacity $z_{\rm mc}(v/u) > 0$; our numerical results confirm this theoretical expectation but find that $z_{\rm mc}(v/u) \to 0$ very rapidly as $v/u \to \infty$. Our numerical results also confirm the theoretical expectation that the corresponding multicritical behavior is in the universality class of the four-state Potts multicritical point on critical line of the two-dimensional Ashkin-Teller model.

\end{abstract}

\maketitle

\section{Introduction}\label{sec:intro}
Dimer models provide interesting examples of entropy-dominated physics~\cite{Henley_Coulombphases}. On planar graphs in the fully-packed limit (i.e., with hole fugacity set to zero), they are exactly solvable by Pfaffian methods\cite{Kasteleyn1961,Temperley1961,Fisher1961}. These methods allow for a detailed characterization of the critical power law correlations of such fully packed dimer models on the square and honeycomb lattices~\cite{Fisher1963}. This critical behavior also admits an interesting description in terms of a coarse-grained action for a fluctuating height field~\cite{Henley_Coulombphases}. 

The analogous fully packed dimer model in three dimensions as well as two-dimensional bilayer models are not exactly solvable even at full packing, nor are related models with an admixture of hard squares. Several such models have been studied using numerical simulations and coarse-grained effective field theory ideas~\cite{Huse_Krauth_Moessner_Sondhi,Ramola2015,Desai2021,Powell2021}. Connections to the physics of quantum dimer models~\cite{Rokhsar1988} and resonating valence bond wave functions~\cite{Albuquerque_Alet,Tang_Sandvik_Henley,Albuquerque_Alet_Moessner} have also been explored~\cite{Damle_Dhar_Ramola,Patil_Dasgupta_Damle}.

Interactions and nonzero hole fugacity also preclude the possibility of an exact solution even in the simple square lattice case. Nevertheless, the phase diagram on the square lattice in the presence of nonzero hole fugacity $z$ and an aligning interaction $u$ that favors two parallel dimers on a square plaquette has been studied in detail using numerical simulations~\cite{Alet2005,Alet2006,Papanikolaou2006}. These studies reveal that the aligning interactions drive a transition to columnar order at low temperature $T$ and fugacity $z$.
In the columnar ordered phase, both lattice rotational symmetry and translational symmetry are spontaneously broken, and almost all dimers align in the same direction.

In this system, the transition from this columnar ordered state to  the dilute high-temperature dimer fluid has a continuously varying correlation length exponent $\nu$, although the anomalous exponent associated with the columnar order parameter remains fixed at $\eta =1/4$ as long as the transition remains second-order in nature~\cite{Alet2006,Papanikolaou2006}. For temperatures below a tricritical value, the transition turns first order~\cite{Alet2006,Papanikolaou2006}. In the regime with a continuously varying correlation length exponent $\nu$, long distance properties are controlled by the physics of the Ashkin-Teller fixed line~\cite{Ashkin1943,Kadanoff1979,Delfino2004}, for which the value of $\nu$ serves as a convenient universal coordinate~\cite{Ramola2015}. Indeed, this particular microscopic realization of Ashkin-Teller criticality is described by the portion of the Ashkin-Teller fixed line that starts at its Kosterlitz-Thouless endpoint (corresponding to the transition in the fully packed dimer model, with $\nu$ formally equal to infinity) and continues on to the four-state Potts point (corresponding to the tricritical transition of this dimer model, with $\nu = 2/3$) on this fixed line~\cite{Alet2006,Papanikolaou2006}.

In related work~\cite{Papanikolaou2014}, Papanikolaou {\em et al} also studied the effect of an additional dimer interaction $v$ that competes with the aligning interaction $u$ and hole fugacity $z$  on the square lattice. The additional interaction represents an attraction between two adjacent dimers on the same principal axis of the square lattice and favors nematic order. In such a nematic state, lattice translation symmetry is preserved, but the symmetry of rotations by $\pi/2$ is spontaneously broken. As a result, $\langle (N_h -N_v)^2 \rangle \sim L^4$ in the thermodynamic limit, where $N_h$ is the number of horizontal dimers, $N_v$ is the number of vertical dimers, and the angular brackets denote the equilibrium average. The presence of such a state at low enough temperature and small enough hole fugacity $z$ was also rigorously established~\cite{Heilmann1979,Jauslin_Lieb2018}.

Consequently, the phase diagram in the presence of both interactions $u$ and $v$ is rich, and supports three different phases at nonzero hole density: a dilute fluid phase, a nematic phase, and a columnar ordered phase.
Previous work~\cite{Papanikolaou2014} characterized the phase diagram in the $T$-$z$ plane in some detail. This analysis led to the following conclusion~\cite{Papanikolaou2014}: When both interactions are nonzero and compete with each other, the transition from the low $z$ low $T$ columnar solid to the fluid proceeds in two steps, an Ising transition from columnar order to nematic order, and a second Ising transition from nematic order to fluid. In this scenario~\cite{Papanikolaou2014} for the phase diagram in the $z$-$T$ plane, there are thus two Ising lines emerging from the $z=0$ Kosterlitz-Thouless transition point when $v$ and $u$ compete with each other. In the $v/u \to \infty$ limit, the first of these pivots coincides with the $z=0$ temperature axis of the $z$-$T$ phase diagram, rendering the low-emperature columnar order unstable to infinitesimal $z$.

Having two Ising transition lines emanate from the Kosterlitz-Thouless transition point on the $z=0$ temperature axis throws up an interesting puzzle when considered from the point of view the coarse-grained field theory ideas used earlier in closely related contexts~\cite{Ramola2015,Papanikolaou2006,Alet2006}. At issue is the fact that the Kosterlitz-Thouless point on the $z=0$ axis at $T=T_{\rm KT}$ is expected to continue on to an Ashkin-Teller line $T_{\rm AT}(z)$ as one turns on a small $z$ and lowers the temperature slightly. 

This fits in with the fact that Kosterlitz-Thouless criticality is known to emerge as the limiting behavior at one end of the Ashkin-Teller line when $\nu \to \infty$ as this end point is approached. Additionally, it is also well-known that if a line of Ashkin-Teller transitions bifurcates into two Ising lines at a multicritical point, this multicritical point is expected to have  four-state Potts symmetry. As a result, one expects that the correlation length exponent tends to $\nu=2/3$ as this point is approached along the Ashkin-Teller line~\cite{Kadanoff1979,AlcarazKoberle1980,Delfino2004}.

Having two Ising transition lines emanate from the Kosterlitz-Thouless transition point on the $z=0$ temperature axis would violate both these expectations. Resolving this puzzle is our principal motivation for revisiting this phase diagram with a pair of complementary techniques, namely, tensor network (TN) computations and large-scale Monte Carlo (MC) simulations using our grand-canonical generalization of the dimer worm algorithm~\cite{Alet2006,Sandvik2006,Syljuasen2002}. 

The tensor network computations use a matrix product operator representation of the row-to-row transfer matrix to obtain a variational uniform matrix product (uMPS) approximation to its top eigenvector (with largest eigenvalue)~\cite{Haegeman2017,ZaunerStauber2018,Vanderstraeten2018,Nietner2020}. This computational method allows efficient scans of large swathes of the phase diagram as well as direct determination of the central charge and scaling dimensions at critical points. Since it works directly in the thermodynamic limit, the accuracy is only limited by the systematic error associated with the finite internal bond dimension for the tensors used in the matrix product representation; this error can be rendered negligible by choosing a large enough bond dimension. In contrast, the MC results only have statistical sampling errors, but include finite-size effects. 
We comment on the conventional transfer matrix method on a stripe.
Although it can be applied to the monomer-dimer model, the width of a stripe is limited because the size of the transfer matrix grows exponentially~\cite{Alet2005,Alet2006}.
Since the maximum width is much smaller than the correlation length obtained by our simulations near the critical point, we do not adopt this method.

In the next section, we define the monomer-dimer model.
In Sec.~\ref{sec:methods}, we explain the numerical methods used in this paper.
Our numerical results are shown in Sec.~\ref{sec:results}.
The last section is devoted to discussion and conclusions.

\begin{figure}[t]
  \centering
  \includegraphics{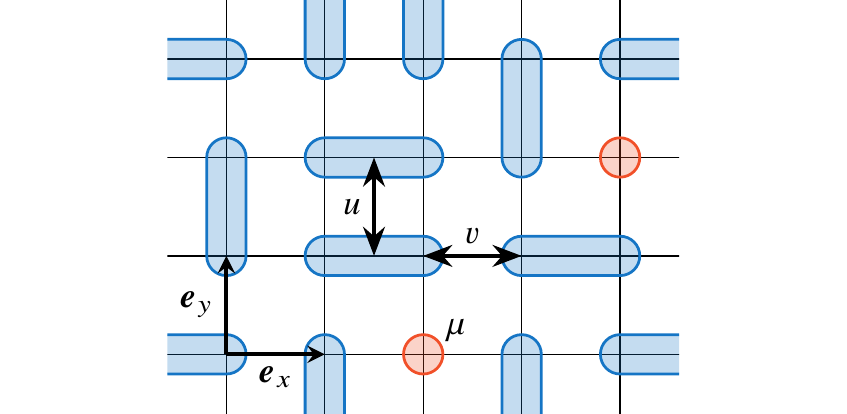}
  \caption{A typical configuration of dimers and monomers.
  Our model has two kinds of dimer-dimer attractive interactions, $u$ and $v$.
  A monomer has the chemical potential $\mu$.}
  \label{fig:config}
\end{figure}

\section{Monomer-Dimer Model}\label{sec:model}

We consider a classical hard-core monomer-dimer model with two kinds of attractive dimer-dimer interactions on the square lattice.
The classical Hamiltonian in the grand canonical ensemble is given as
\begin{multline}
  H = - \sum_{\bm{r}}\sum_{\alpha=x, y} \bigl\{
    u\, n_\alpha(\bm{r})\, n_\alpha(\bm{r}+\bm{e}_{\beta\neq\alpha}) \\
    + v\, n_\alpha(\bm{r})\, n_\alpha(\bm{r}+2 \bm{e}_{\alpha})\bigr\}
    - \mu \sum_{\bm{r}} \, n_\text{m}(\bm{r}).
    \label{eq:H}
\end{multline}
Here, $n_\alpha(\bm{r})$ denotes the dimer occupation number of a link between neighboring sites, $\bm{r}$ and $\bm{r}+\bm{e}_\alpha$,
and $n_\text{m}(\bm{r})$ is the monomer occupation number at a site $\bm{r}$.
These occupation numbers take the value $0$ or $1$, and should satisfy $n_x(\bm{r}) + n_y(\bm{r}) + n_x(\bm{r}-\bm{e}_x) + n_y(\bm{r}-\bm{e}_y) + n_\text{m}(\bm{r}) = 1$ because of the hard-core constraint.

\begin{figure*}[t]
  \centering
  \includegraphics{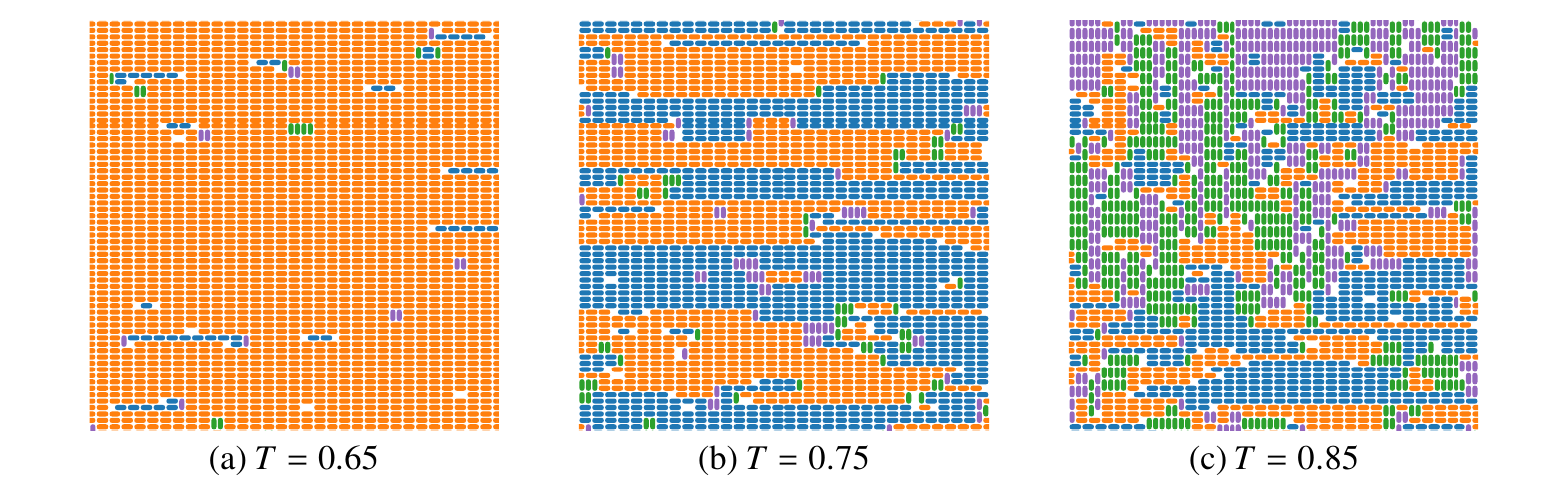}
  \caption{A typical monomer-dimer configuration of the (a) columnar ordered, (b) nematic, and (c) disordered fluid phases. These are a part of snapshots generated by the MC simulations of the $L=256$ system at $(u,v,z)=(0.1,0.9,0.2)$.
  A color of dimers corresponds to a value of the local order parameter Eqs.~\eqref{eq:psi_col} and \eqref{eq:psi_nem}.
  Blue, orange, green and purple dimers have $(\Psi_\text{col}(\bm{r}), \Psi_\text{nem}(\bm{r}))=(1, 1)$, $(-1, 1)$, $(i, -1)$ and $(-i, -1)$, respectively.
  Monomers are indicated by a empty site.}
  \label{fig:snapshot}
\end{figure*}

The $u$ term in the first term of Eq.~\eqref{eq:H} is the interaction between two dimers on a plaquette.
On the other hand, the $v$ term acts on two adjacent dimers aligning in the dimer direction (Fig.~\ref{fig:config}).
We call the former the plaquette interaction and the latter the dimer-aligning interaction as well as Ref.~\cite{Papanikolaou2014}.
We assume that $u$ and $v$ are non-negative, that is, both the interactions are attractive.
The last term of Eq.~\eqref{eq:H} is the chemical potential of monomers.
The fugacity of a monomer at the temperature $T$ is defined as $z\equiv e^{\beta \mu}$, where $\beta=1/T$ is the inverse temperature.
In this paper, we set $u+v=1$ as a unit of energy so that the perfectly columnar ordered state at full packing has a constant energy per site, $e=-(u+v)/2$.

When the plaquette interaction $u$ is sufficiently large, the columnar ordered phase appears.
This phase spontaneously breaks the symmetry of lattice rotations about a site, as well as lattice translational symmetry along one principal axis.
In contrast, in the nematic phase which is favored by the $v$ term, the system spontaneously choses to have macroscopically more dimers of one orientation over the other. Translational symmetries along both principal axes are preserved, but the system spontaneously breaks the symmetry of lattice rotations about a site. 

Typical configurations of three phases are shown in Fig.~\ref{fig:snapshot}.
These are generated by the MC simulations at $v=0.9$ and $z=0.2$, using the method described in the next section for three temperatures, $T=0.65$, $0.75$, and $0.85$, which  correspond respectively to the columnar ordered, nematic and disordered fluid phases. From this depiction, it is clear that the nematic phase breaks the lattice rotational symmetry but does not break translational symmetry, while the columnar state breaks both lattice translation symmetry and rotational symmetry.

\section{Methods}\label{sec:methods}
As mentioned in Introduction, our computational study of this monomer-dimer model uses complementary methods.
One employs our grand-canonical generalization of the dimer worm algorithm to perform Monte Carlo simulations, while the other uses tensor network methods. Below, we summarize each in turn.

\subsection{Monte Carlo method}\label{sec:MC}
The usual dimer worm algorithm~\cite{Syljuasen2002,Sandvik2006,Alet2006} provides a rejection-free nonlocal update scheme for interacting dimer models at full-packing. Here, we build on ideas developed in Ref.~\onlinecite{Rakala_Damle} to generalize this dimer worm algorithm and obtain an efficient grand-canonical algorithm for the monomer-dimer model at nonzero monomer fugacity.
In the first step of our grand-canonical scheme, one chooses at random a site $j_{\rm init}$ of the lattice.
There are two possibilities at this first step: either the initially chosen site $j_{\rm init}$ has a monomer on it, or it is covered by a dimer. Let us consider each in turn.

\begin{figure}[t]
  \centering
  \includegraphics{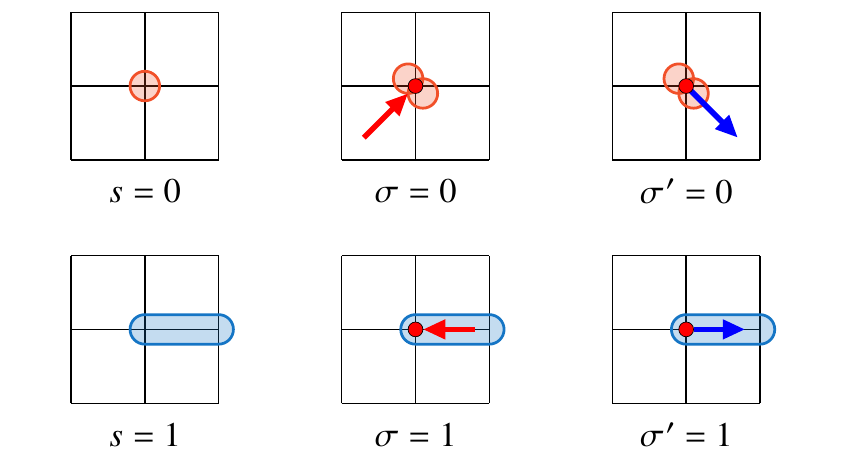}
  \caption{The label $s$ labels the different allowed states of a site in our monomer-dimer model. Other dimer states ($s=2, 3, 4$) not shown here are generated by the counter-clockwise $90^{\circ}$ rotations.
  The label $\sigma$ ($\sigma'$) labels the different entrance (exit) configurations corresponding to a pivot, as defined in our grand-canonical worm algorithm. In the latter context of the worm algorithm, the configurations shown provide a pictorial illustration of the local information needed for calculating the reduced weights given in Eq.~\eqref{eq:Weights}, which enter the detailed balance constraint equations Eq.~\eqref{eq:ProbTable}. Thus entrance/exit $\sigma =0$ is schematically depicted with {\em two} monomers at the pivot in question to emphasize that the corresponding reduced weight is $z^2$, not $z$. This is appropriate since entrance/exit $\sigma = 0$ ($\sigma'=0$) corresponds to a configuration that has one less {\em dimer} compared to the configurations associated with other entrances/exits.
  }
  \label{fig:states}
\end{figure}

If $j_{\rm init}$ has a monomer on it, we have five options at our disposal. The first four options consist of placing a dimer connecting the initial site to one of its four neighbors. The fifth option is to exit without doing anything.  Each of these possibilities is assigned a probability from a probability table. We will discuss the construction of this probability table in some detail below. For now we simply introduce some language that will subsequently be useful in describing the construction of this probability table: the initial site is our first ``pivot'' site $\pi_0$, which we have ``entered'' from the ``entry'' $\sigma_0 = 0$, {\em i.e.} from ``outside the lattice''  (Fig.~\ref{fig:states}). Aborting our attempted worm move at this step itself without doing anything corresponds to ``exiting the pivot'' $\pi_0$ via ``exit'' $\sigma'_0 = 0$.  On the other hand, if we opt to place a dimer connecting the pivot $\pi_0$ to its $k$-th neighbor, this option corresponds to exiting the pivot via exit $\sigma'_0 = k$ (so $k$ can take on values from 1 to 4). If the chosen exit is $\sigma'_0 \neq 0$ , we now move to the site corresponding to the chosen exit and continue the construction.

Before we describe what is done next, we need to specify the procedure to be used if the initially chosen site $j_{\rm init}$ has a dimer covering it. In this case, one walks to the other end of this dimer; the site covered by this other end becomes our first pivot $\pi_0$, which we have ``entered'' from the entry $\sigma_0$ corresponding to $j_{\rm init}$. Now, the choices available are again five in number: One can delete this dimer that connects the pivot site $\pi_0$ to the entrance site $j_0$. This introduces two monomers in the system and concludes the worm move. As before, this corresponds to ``exiting'' the first pivot $\pi_0$ via exit $\sigma'_0 = 0$. Or, one can pivot the dimer covering $\pi_0$ so that it now connects $\pi_0$ to its $k$-th neighbor. If one of these latter four options is chosen, we say the pivot $\pi_0$ is exited via exit number $\sigma'_0 = k$, and we move to the site corresponding to the chosen exit to proceed further as described below.

At this stage of our worm construction, we are at the site corresponding to exit $\sigma'_0$ of the previous pivot, having arrived there because we chose to place a dimer connecting the previous pivot point $\pi_0$ to this exit site. If this site does not already have another dimer covering it, we have reached an allowed configuration and the worm move ends. On the other hand, if this exit site does have another dimer already covering it, this site becomes the current ``overlap site'' $o_0$. We now walk along this pre-existing dimer from $o_0$ to its other end. The site at this other end becomes our next pivot site $\pi_1$, which has been ``entered'' via entry number $\sigma_1$ that corresponds to the overlap site $o_0$. 

At this step, there are again five choices for $\sigma'_1$, the exit to be used to exit the current pivot site $\pi_1$. As before, exit $\sigma'_1 =0$  corresponds to deleting the dimer covering the current pivot site $\pi_1$. If this is chosen, the worm move ends. On the other hand, exits numbered $\sigma'_1=1$ through $\sigma'_1=4$ correspond to pivoting the dimer covering $\pi_1$ so that it now connects $\pi_1$ to the site corresponding to $\sigma'_1$. If this exit site does not already have a dimer covering it, we have reached an allowed configuration and the move ends. Otherwise, this exit site becomes the next overlap site $o_1$, and the procedure is repeated.

It is easy to see that this worm construction yields a valid rejection-free algorithm if we choose the probability table $P_{\sigma \rightarrow \sigma'}$ for transition probabilities in a way that it satisfies local detailed balance at each step. This amounts to requiring that the probability obeys the constraint equations:
\begin{equation}
  \omega_{\sigma} P_{\sigma\rightarrow\sigma'} =
  \omega_{\sigma'} P_{\sigma' \rightarrow \sigma}.
  \label{eq:ProbTable}
\end{equation}
Here, $P_{\sigma \rightarrow \sigma'}$ is the conditional probability for exiting a pivot via exit $\sigma'$ given that we have entered it via entrance $\sigma$, $P_{\sigma' \rightarrow \sigma}$ is the conditional probability for the reverse process, and the weights $\omega_{\sigma}$ and $\omega_{\sigma'}$ represent the Boltzmann weights of the configurations corresponding to the choices $\sigma$ and $\sigma'$, respectively. These Boltzmann weights are to be calculated ignoring the violation of the hard-core constraint on dimers in the configurations that arise during the worm construction.

The simplest choice of solution is the heat-bath solution (sometimes called the Gibbs sampler) given as $P_{\sigma\rightarrow\sigma'} = \omega_{\sigma'} / \sum_{\sigma'}\omega_{\sigma'}$. In practice, we use the iterative Metropolized Gibbs sampler to reduce the bounce process~\cite{Pollet2004}, i.e., reduce the magnitude of the diagonal elements of the probability table. Note also that the computation of the weights $\omega_{\sigma}$ and $\omega_{\sigma'}$ is simplified by the fact that they only differ due to factors arising from the contribution of the immediate neighborhood of the pivot. Since the equation set is homogenous, we can cancel all common factors to define reduced weights that only depend on the local environment of the pivot, and use these in Eq.~\eqref{eq:ProbTable}.

These reduced weights can be written as
\begin{equation}
  \omega_{\sigma} = 
    z^2\delta_{\sigma, 0}
    + e^{\beta(un+vm)}(1-\delta_{\sigma, 0})
    \label{eq:Weights}
\end{equation}
where $n$ ($m$) denotes the number of nearest neighbor dimers parallel in the transverse (longitudinal) direction to the dimer that covers the pivot when the configuration corresponds to entrances/exits $\sigma \neq 0$.
These numbers $n, m \in \{0, 1, 2\}$ can be calculated by checking the direction of dimers on eight sites around $\bm{r}_\text{h}$, that is, 
$\bm{r}_\text{h} \pm \bm{e}_x$,
$\bm{r}_\text{h} \pm \bm{e}_y$,
$\bm{r}_\text{h} \pm 2\bm{e}_x$,
$\bm{r}_\text{h} \pm 2\bm{e}_y$.
The number of valid configurations in these eight sites is $65089$, much smaller than $5^8=390625$.

The factor of $z^2$ in the first term of Eq.~\eqref{eq:Weights} reflects the fact that configuration $\sigma = 0$ has one fewer dimer, i.e., two additional holes (monomers) in comparison with the configurations with $\sigma \neq 0$.

Finally, we note for completeness that this worm construction and its detailed balance property generalizes straightforwardly to lattices with arbitrary coordination number. The ``out-of-plane'' entrance/exit in the general case is numbered $0$, and the other entrances/exits are numbered from $1$ to $n_c$, where $n_c$ is the coordination number of the pivot site in question ($n_c$ can be different for different sites, and there is thus no restriction of regularity for this algorithm to remain valid)

Using this algorithm, we perform the MC simulations of $N=L\times L$ systems with periodic boundary conditions for system size $L$ up to $L=512$. The number of the worm updates $n_w$ used per Monte Carlo step is chosen for each set of control parameters to be such that $n_w \langle l_w \rangle = N$, $\langle l_w \rangle$ is the mean number of sites visited during the construction of a single worm.
We perform $10^3\times N$ worm updates to estimate $\langle l_w \rangle$ and to thermalize the system before measuring physical quantities.
With this convention defining a MC step, we ensure that we obtain at least $2 \times 10^6$ MC configurations of the system from which we can calculate equilibrium properties.

\subsection{Tensor network method}\label{sec:TN}

The tensor network representation of our model is based on the singular value decomposition of the local Boltzmann weight on a bond.
The partition function is rewritten as the contraction of the tensor network,
\begin{equation}
  Z = \text{tTr} \bigotimes_{i} A.
\end{equation}
The tensor $A$ is located on sites of the square lattice and has four indices representing links to the nearest neighbor sites.
An element of $A$ has the form
\begin{equation}
  A_{xyx'y'} = \sum_{s=0}^{4}
    (X_r)_{sx} (X_l)_{sx'}
    (X_t)_{sy} (X_b)_{sy'}
    Y_s,
\end{equation}
where $s$ denotes the local configuration at a site as shown in Fig.~\ref{fig:states}.
The matrices $X$'s are determined by the singular value decomposition of the local Boltzmann weight on a bond.
The Boltzmann weight on a horizontal bond is represented as a $5\times 5$ matrix,
\begin{equation}
  W_h = 
  \begin{pmatrix}
    1 & 1 & 1 & 0 & 1 \\
    0 & 0 & 0 & 1 & 0 \\
    1 & 1 & e^{\beta u/2} & 0 & 1 \\
    1 & e^{\beta v} & 1 & 0 & 1 \\
    1 & 1 & 1 & 0 & e^{\beta u/2} \\
  \end{pmatrix},
\end{equation}
where the row (column) index of $W_h$ corresponds to a state at $\bm{r}$ ($\bm{r}+\bm{e}_x$), respectively.
Elements with a value of zero indicate a configuration prohibited by the hard-core constraint.
The singular value decomposition, $W_h = U_h S_h V_h^{T}$, defines $X_r\equiv U_h S_h^{1/2}$ and $X_l\equiv V_h S_h^{1/2}$.
We note that $U_h$ and $V_h$ can be chosen to be real orthogonal $5\times 5$ matrices since $W_h$ is a real square matrix.
Similarly we obtain the Boltzmann weight on a vertical bond 
between $\bm{r}$ (row) and $\bm{r}+\bm{e}_y$ (column) as
\begin{equation}
  W_v =
  \begin{pmatrix}
    1 & 1 & 1 & 1 & 0 \\
    1 & e^{\beta u/2} & 1 & 1 & 0 \\
    0 & 0 & 0 & 0 & 1 \\
    1 & 1 & 1 & e^{\beta u/2} & 0 \\
    1 & 1 & e^{\beta v} & 1 & 0 \\
  \end{pmatrix}
  = U_v S_v V_v^{T},  
\end{equation}
and we define $X_t \equiv U_v S_v^{1/2}$ and $X_b \equiv V_v S_v^{1/2}$.
The chemical potential of a monomer acts as the external field and gives the corresponding on-site factor as
\begin{equation}
  Y_s = z\, \delta_{s, 0} + (1-\delta_{s, 0}).
\end{equation}
The first term has the factor of $z$ in contrast to Eq.~\eqref{eq:Weights}.
The former corresponds to the Boltzmann weight for valid configurations of our monomer-dimer model, while tha latter is for extended configurations containing one doubly occupied site.

The row-to-row transfer matrix with infinite width is represented as a uniform matrix product operator with a local tensor $A$.
Using the variational uniform matrix product state algorithm (VUMPS)~\cite{Haegeman2017,ZaunerStauber2018,Vanderstraeten2018,Nietner2020}, we calculate the eigenvector corresponding to the largest eigenvalue of the transfer matrix.
The uniform matrix product state (uMPS) obtained in this way approximates this eigenvector with accuracy that is controlled by the bond dimension $\chi$ of the uMPS.
We increase $\chi$ up to $128$ to ensure sufficient accuracy. In practice, we assume that the uMPS has a $2\times 2$ unit-cell structure~\cite{Nietner2020} as is appropriate for a description of  the columnar ordered state.
After calculating the horizontal uMPS in this way, we also calculate the vertical uMPS, which approximates the corresponding eigenvector of the column-to-column transfer matrix.
A good initial guess for the vertical uMPS can be given by the fixed point tensor of the horizontal uMPS~\cite{Haegeman2017}.
We have confirmed that the physical quantities calculated from the horizontal and vertical uMPS agree with each other to machine precision even in the ordered phases.

\section{Observables and interpretation}
We now summarize the definitions and physical significance of the various observables of interest to us in this problem, and indicate how they may be accessed in either of the computational methods we use.

\begin{figure}[t]
  \centering
  \includegraphics{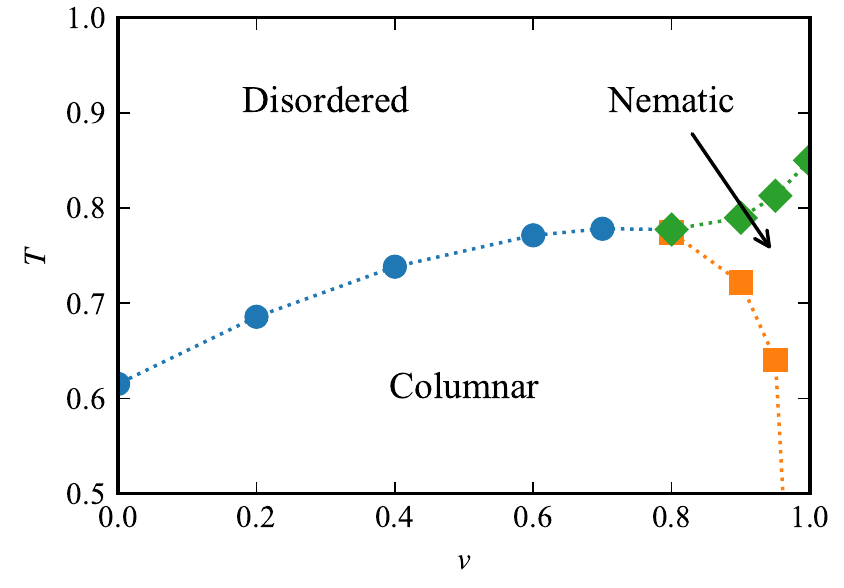}
  \caption{Phase diagram at $z=0.2$ and $u+v=1$ calculated by the TN simulation. The nematic phase exists between diamond and square symbols. The transition temperature between the columnar and nematic phases becomes zero at $v=1$. The finite bond-dimension effect is smaller than the symbol size.}
  \label{fig:phase_diagram}
\end{figure}

\subsection{The order parameters and Binder ratios}

We detect columnar order using a complex order parameter constructed from the following local order parameter field defined at each site $\bm{r}=(r_x, r_y)$ as
\begin{multline}
  \Psi_\text{col}(\bm{r}) \equiv (-1)^{r_x}
  \left\{ n_x(\bm{r}) - n_x(\bm{r}-\bm{e}_x) \right\}\\
  + i\,(-1)^{r_y} 
  \left\{ n_y(\bm{r}) - n_y(\bm{r}-\bm{e}_y) \right\}.
  \label{eq:psi_col}
\end{multline}
The corresponding order parameter,
\begin{equation}
  m_\text{col} \equiv \frac{1}{N} \sum_{\bm {r}}
    \Psi_\text{col}(\bm{r}),
\end{equation}
takes $\pm 1$ or $\pm i$ when the state has the complete columnar order.

Nematic order, which breaks the symmetry of $\pi/2$ rotations, can be detected by comparing the number of horizontal and vertical dimers. With this motivation, we define the local nematic order parameter field as
\begin{equation}
  \Psi_\text{nem}(\bm{r}) \equiv n_x(\bm{r}) + n_x(\bm{r}-\bm{e}_x)
  - n_y(\bm{r}) - n_y(\bm{r}-\bm{e}_y),
  \label{eq:psi_nem}
\end{equation}
The corresponding order parameter,
\begin{equation}
  m_\text{nem} \equiv \frac{1}{N} \sum_{\bm{r}} \Psi_\text{nem}(\bm{r})
  = \frac{2}{N} \sum_{\bm{r}} \left\{ n_x(\bm{r}) - n_y(\bm{r})\right\} ,
\end{equation}
takes on values $\pm 1$ both in the nematic and columnar states.

The corresponding Binder ratios~\cite{Binder1981} are defined in the usual way:
\begin{equation}
  U_\text{col} \equiv
  \frac{\left< |m_\text{col}|^4 \right>}{\left< |m_\text{col}|^2 \right>^2}, \quad
  U_\text{nem} \equiv
  \frac{\left< m_\text{nem}^4 \right>}{\left< m_\text{nem}^2 \right>^2}.
\end{equation}
As is well-known, the Binder ratios converge in the ordered phase to $1$  as the denominator and numerator take on the same limiting value in the thermodynamic limit. On the other hand, in a phase without symmetry breaking, the limiting value depends on the nature of fluctuations of the order parameter.
$U_\text{col}$ in the nematic and $U_\text{nem}$ in the disordered fluid phase are both expected to converge to $3$ in the thermodynamic limit because the fluctuations of the corresponding order parameters obey a one-dimensional Gaussian distribution in these regimes. 
On the other hand, $U_\text{col}\rightarrow 2$ in the disordered fluid phase because the fluctuations of $m_\text{col}$ obey a two-dimensional Gaussian distribution.
We note that the conventional and Ising definition of the Binder parameter for the frustrated Ising model discussed in Ref.~\cite{Watanabe2021} correspond $U_\text{col}$ and $U_\text{nem}$, respectively.
Since the Binder ratio is a dimensionless quantity in the sense of the renormalization group, curves that represent the dependence of a Binder ratio on a control parameter ($z$ or $T$) for systems of different sizes $L$ are all expected to cross at the critical value of the control parameter. This allow us to locate the phase transitions involving loss of order in a convenient way.

\begin{figure}
  \centering
  \includegraphics{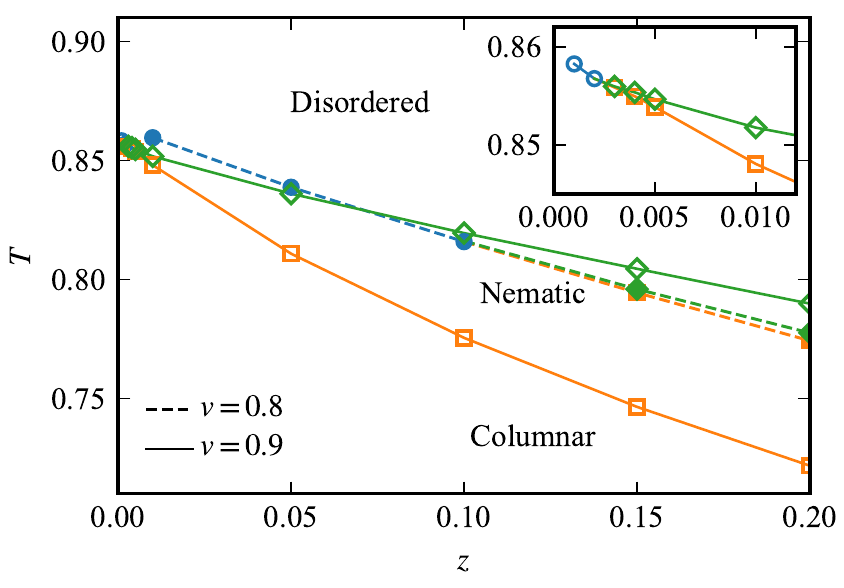}
  \caption{
    Phase diagram on the $T$--$z$ plane obtained by the TN simulation with $\chi=128$.
    The open (filled) symbols with the solid (dashed) line indicate transition points at $v=0.9$ ($0.8$) with $u=1-v$.
    The direct phase transition between the columnar ordered and disordered fluid phases is shown by a blue circle as well as Fig.~\ref{fig:phase_diagram}.
    The inset is a magnified view for $v=0.9$.
  }
  \label{fig:phase_zT}
\end{figure}

\subsection{Correlation length and entanglement entropy}

The correlation length is obtained as
\begin{equation}
  \xi = - \frac{2}{\ln (\lambda_1 / \lambda_0)}
  \label{eq:col_length}
\end{equation}
where $\lambda_0$ ($\lambda_1$) is the (second) largest eigenvalue of the transfer matrix defined by the uMPS.
We note that the uMPS is always normalized such that $\lambda_0=1$, and 
the factor $2$ comes from the unit-cell size of the uMPS.
Since we calculate the uMPS in both the horizontal and vertical directions, two kinds of correlation length exist in a phase that breaks the symmetry of lattice rotations.
The ``transverse'' correlation length $\xi_\perp$ is measured along the direction perpendicular to the dimers, while the ``longitudinal'' correlation length $\xi_\parallel$ is in the direction parallel to the dimers.
In the disordered fluid phase, both correlation lengths are the same as expected.
In the nematic phase, a level crossing of $\lambda_1$ occurs, and $\xi_\perp$ takes a smaller value than $\xi_\parallel$ below a certain temperature.
In the columnar ordered phase, we always have $\xi_\perp < \xi_\parallel$.
In other words, the correlation length scale along the dimers become larger than in the perpendicular direction.
Thus, the transverse correlation length seems to be suitable for studying the phase transition between the columnar ordered and disordered fluid phases.
We also note that $\xi$ corresponds to the truncated correlation function and takes a finite value even in the ordered phase.

The entanglement entropy is defined as
\begin{equation}
  S_\text{EE} = - \sum_{i} \sigma_i^2 \ln \sigma_i^2,
\end{equation}
where $\sigma_i$ denotes the singular value of the core matrix in the mixed canonical form of the uMPS.
Since we calculate the horizontal and vertical uMPS with a $2\times 2$ unit-cell structure, $S_\text{EE}$ may depend on direction and position in the unit cell. In the disordered and nematic phases, we find that the all $S_\text{EE}$ are equal to each other. On the other hand, in the columnar ordered phase, we find that $S_\text{EE}$ takes on two values, depending on whether the core matrix of the horizontal uMPS is on a dimer or between dimers.
The former always yields the larger $S_\text{EE}$.
In our analysis below, we use the largest entanglement entropy thus obtained.

\subsection{Connection with coarse-grained Ashkin-Teller description}
It is instructive to think in terms of a coarse-grained version $\psi$ of our local complex columnar order parameter field $\Psi_\text{col}$, and write
\begin{eqnarray}
\psi &\propto& (\tau_1 + \tau_2) + i(\tau_1 - \tau_2)
\end{eqnarray}
Clearly, the corresponding coarse-grained version $\phi$ of the local nematic order parameter field $\Psi_\text{nem}$ satisfies
\begin{eqnarray}
\phi &\propto & \text{Re} ( \psi^2) \nonumber \\
&\propto & \tau_1 \tau_2
\end{eqnarray}

The $\tau$ defined in the above are two coarse-grained Ising fields. In the columnar ordered phase, both $\tau_1$ and $\tau_2$ are ordered; this correctly accounts for the four-fold symmetry breaking in the columnar ordered state. Nematic order corresponds to the product $\tau_1 \tau_2$ being ordered, without any long range order in the individual $\tau$. From the symmetries of the original problem, we see that interchanging the $\tau$ is a symmetry of the theory. Thus, the natural description is in terms of a symmetric Ashkin-Teller theory with two Ising fields $\tau_1$ and $\tau_2$.

This connection to the physics of the Ashkin-Teller model~\cite{Ashkin1943,Kadanoff1979,Delfino2004} yields a wealth of information. For instance, along a line of continuous transitions from the columnar ordered state to the disordered fluid state, we expect the critical behavior to controlled by the critical properties of the corresponding fixed line in the Ashkin-Teller model. Along this line, both the Ising fields $\tau_1$ and $\tau_2$ have a fixed anomalous dimension of $\eta =1/4$~\cite{Ashkin1943,Kadanoff1979,Delfino2004}. Since the columnar order parameter is linear in the Ising fields $\tau_{1/2}$, we expect it to also scale with an anomalous exponent $\eta =1/4$ all along the line of continuous transitions between columnar ordered and disordered fluid phases. 

Along the fixed line of the Ashkin-Teller theory, $\tau_1 \tau_2$ scales with an anomalous dimension $\eta_2$ that varies continuously~\cite{Ashkin1943,Kadanoff1979,Delfino2004} and is related to the continuously varying correlation length exponent by the Ashkin-Teller relation $\eta_2 = 1 - 1/2 \nu$. Since the nematic order parameter $\phi \sim \tau_1 \tau_2$, we expect it to have an anomalous dimension $\eta_2$~\cite{Ramola2015} given by this relation all along the line of continuous transitions between columnar ordered and disordered fluid phases. 

In the Ashkin-Teller model, the point at which Ashkin-Teller line splits into two lines of Ising transitions is known to have the symmetries for the four-state Potts model~\cite{Ashkin1943,Kadanoff1979,Delfino2004,AlcarazKoberle1980} ; in the phase between these two Ising transition lines, $\tau_1 \tau_2$ is ordered although $\tau_1$ and $\tau_2$ remain individually disordered . The enhanced Potts symmetry at this multicritical point implies that $\tau_1$, $\tau_2$, and $\tau_1 \tau_2$ all have the same anomalous exponent. Thus $\eta_2 = \eta = 1/4$ at this point, and the Ashkin-Teller relation implies $\nu = 2/3$. Given the correspondence made above, this implies that the nematic order parameter is expected to have an anomalous exponent of $1/4$ at the multicritical point at which the Ising phase boundaries of the nematic phase meet the line of continuous transitions between columnar ordered and disordered fluid phases.

From this perspective, it is clear that the value of $\nu$ (or equivalently $\eta_2$) serves as a universal coordinate for the line of continuous transitions from the columnar ordered phase to the  disordered fluid phase. At full-packing, i.e., $z= 0$, the system has a description in terms of a coarse-grained Gaussian height action for a scalar height $h$, and the transition from the power-law ordered high temperature state to the low temperature state is expected to be governed by a Kosterlitz-Thouless transition at which the leading cosine nonlinearity $\cos(8 \pi h)$ becomes relevant. As a result, one expects $\nu \to \infty$ as $z \to 0$ along the line of continuous transitions between the columnar ordered state and the disordered fluid state. From this it is clear that the multicritical point at which the two Ising lines meet cannot be at $z=0$, since this multicritical point corresponds to a value of $\nu = 2/3$.

This theoretical perspective and the resulting expectations informs much of the data analysis we present in the next section.

\section{Numerical results}\label{sec:results}

Before getting into the details, it is useful to provide a summary of our results for representative slices through the phase diagram, as these slices clarify the overall picture and help answer the question raised in Introduction.

\subsection{Overview}

To this end, we first consider a fixed $z=0.2$ slice and display the computed two-dimensional phase diagram in the $T$--$v$ plane (with $u=1-v$). This is shown in Fig.~\ref{fig:phase_diagram}. 

For $v\leq 0.7$, there is a direct temperature driven phase transition between the columnar ordered and disordered fluid phases. As $v$ is increased further, this phase boundary splits slightly below $v=0.8$ into two transition lines and the nematic phase appears as an intermediate phase beyond this multicritical point at which three transition lines meet. We have also checked that a corresponding slice at somewhat larger $z$ reduces the temperature scale of the transitions.  The transition temperatures shown in Fig.~\ref{fig:phase_diagram} is determined from the peak position of the correlation length estimated by the TN method.
The result of the MC method agrees with it within errors that are smaller than the symbol sizes used.

Next we consider slices with fixed $v=0.9$ and $v=0.8$ (with $u=1-v$) and display the computed two-dimensional phase diagram in the $T$--$z$ plane. This is shown in Fig.~\ref{fig:phase_zT}.
The transition points are determined in the same way as Fig.~\ref{fig:phase_diagram}. Since the fully packed $z=0$ system is particularly challenging for TN computations, we do not extend our study all the way to $z=0$. Nevertheless, we are able to go to low enough $z$ to demarcate the essential features of the phase diagram.
It would be difficult to obtain this phase diagram using the MC method because the phase boundaries cannot be calculated with such precision due to the finite size effect.

At $v=0.8$, the low-temperature nematic phase is quite narrow and disappears below the multicritical value of $z$ which is close to $z=0.1$.
On the other hand, for $v=0.9$, the nematic phase is very broad and seems to exist even at very low values of $z$. However, our detailed computations reveal that there is no nematic state below a multicritical threshold value of $z$ which is close to $z=0.002$ (Fig.~\ref{fig:phase_zT}). The actual monomer densities associated with these multicritical points are extremely small: For $v=0.8$, the multicritical monomer density is about $\delta=0.005$. At $v=0.9$, the multicritical monomer density is a much lower value of $5\times 10^{-5}$.

This conclusion is contrary to that of Ref.~\cite{Papanikolaou2014}. 
However, it is entirely consistent with
our understanding of the phase boundaries based on the coarse-grained effective field theory.
Indeed, as we have already reviewed earlier,  the multicritical point is expected to have rather different universal behavior from the transition at full packing since the former is expected to correspond to a $\eta_2 =1/4$ and the latter corresponds to $\eta_2 = 1$ (where $\eta_2$ is the anomalous exponent associated with the nematic order parameter). As a result, for $v/u < \infty$, there is no consistent scenario in which the multicritical point coincides with the transition at full packing.
The resolution is of course that the multicritical value of $z$ approaches $z=0$ extremely rapidly as $v/u$ is increased, but does not reach $z=0$ at any finite $v/u$. 

In the rest of this section, we display our results for a few representative values of $v$ and $z$, and provide a detailed account of the analysis that leads to these phase diagrams and this overall conclusion. 

\subsection{Detailed analysis}

\begin{figure}
  \centering
  \includegraphics{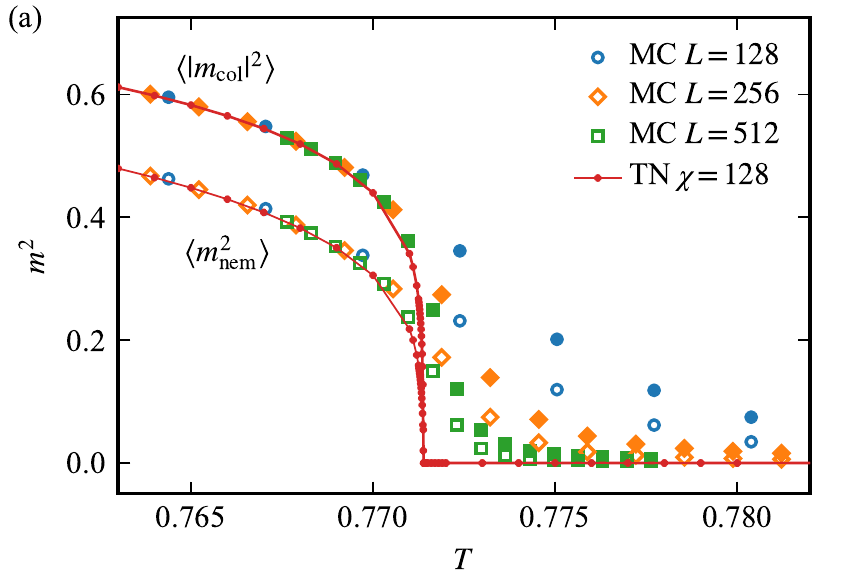}
  \includegraphics{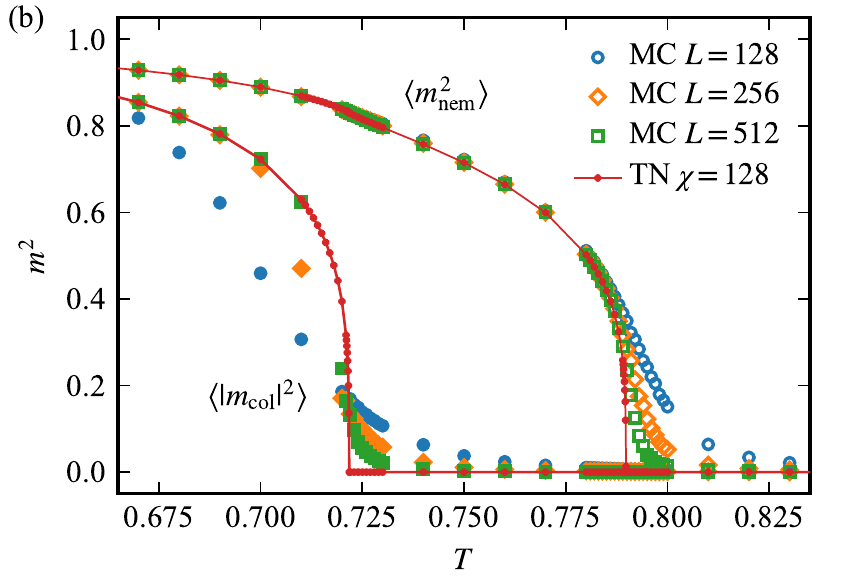}
  \caption{The square of order parameters at (a) $(u,v,z)=(0.4, 0.6, 0.2)$ and (b) $(0.1, 0.9, 0.2)$. The filled (open) symbols denote the columnar (nematic) order parameter. The statistical error is smaller than the symbol size. The results by TN with $\chi=128$ are also shown by the small dots connected by a line.}
  \label{fig:order_param}
\end{figure}

We use both tensor network (TN) and Monte Carlo (MC) methods to obtain the nematic and columnar order parameters, since these two complementary methods provide a nontrivial consistency check on each other.
Figure \ref{fig:order_param} shows the temperature dependence of the order parameters at $v=0.6$ and $v=0.9$ with $z=0.2$. 

\begin{figure}
  \centering
  \includegraphics{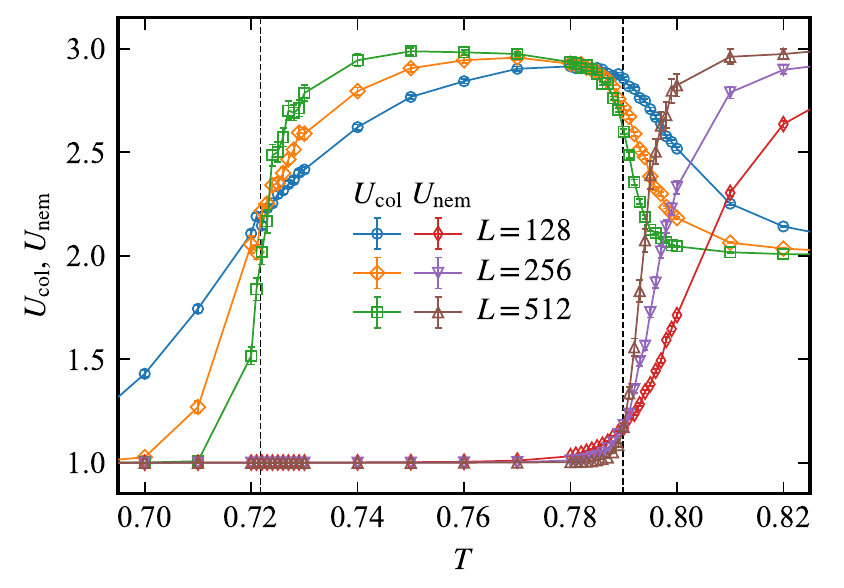}
  \caption{The Binder ratio for columnar and nematic orders at $(u,v,z)=(0.1, 0.9, 0.2)$. The dashed vertical lines denote the critical temperature obtained by the TN method.}
  \label{fig:binder}
\end{figure}

At $v=0.6$, both the order parameters take on nonzero values below $T=0.7714$, signaling the onset of columnar order.
On the other hand, at $v=0.9$, we find that a nematic phase exists between $T_{1}=0.722$ and $T_{2}=0.790$ , as is clear from the fact that the columnar order parameter vanishes but the nematic order parameter takes on a nonzero value. The transition temperatures are estimated by identifying the crossing points of the Binder ratios, as shown in Fig.~\ref{fig:binder} for the Binder ratios calculated by the MC simulations at $v=0.9$ and $z=0.2$. These crossing points are found to be consistent with the transition temperature estimated by the TN method.

We obtain the scaling dimensions of various quantities at these transitions by performing a finite-size scaling (FSS) analysis, using the following FSS form for the columnar and nematic order parameters,
\begin{equation}
  \langle |m_a|^2 \rangle \sim L^{2x_a} f\left((T-T_c) L^{1/\nu}\right),
  \label{eq:fss}
\end{equation}
where $x_a$ denotes the scaling dimension of the corresponding operator $\Psi_a$ ($a=\text{col}, \text{nem}$); these scaling dimensions are related to the anomalous exponents introduced earlier via $x_\text{col} = \eta/2$, $x_\text{nem} = \eta_2/2$.
Likewise, the scaling dimension of the energy operator is related to the correlation length exponent introduced earlier: $x_t = 2- 1/\nu$.

\begin{figure}
  \centering
  \includegraphics{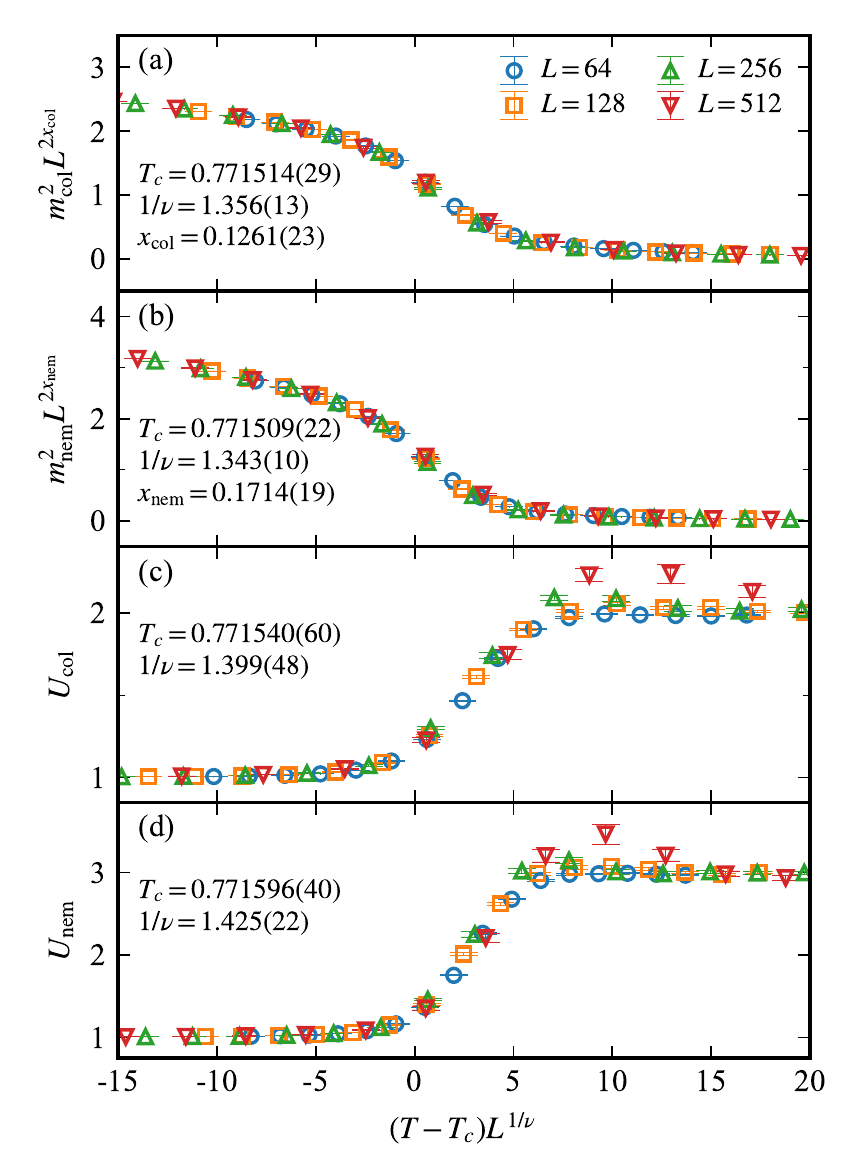}
  \caption{The FSS plots  at $(u, v, z)=(0.4, 0.6, 0.2)$ for (a) the columnar order parameter, (b) the nematic order parameter, (c) the columnar Binder ratio, and (d) the nematic Binder ratio.}
  \label{fig:fss}
\end{figure}

We use the kernel method~\cite{Harada2011} to infer the critical exponents and the critical temperature and estimate their confidential intervals.
At the best fit value of the scaling dimensions and the transition point, all data collapse reasonably well onto a single curve as shown in Fig.~\ref{fig:fss}.

\begin{figure}
  \centering
  \includegraphics{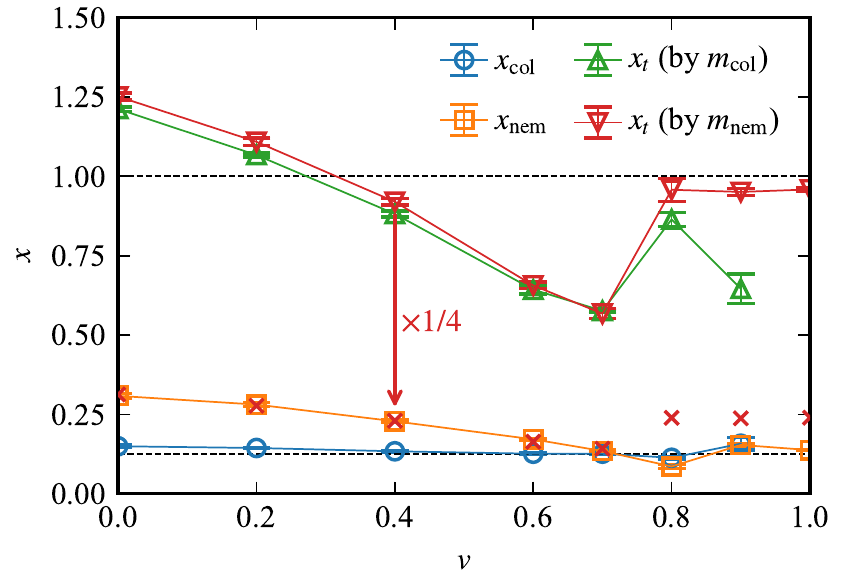}
  
  \caption{The scaling dimensions estimated by the FSS analysis of the order parameters~\eqref{eq:fss} at $z=0.2$. The horizontal dashed lines indicate $x_h=1/8$ and $x_t=1$ in the Ising universality class.
  The red crosses indicate $x_t/4$, which should be equal to $x_{\text{nem}}$ on the Ashkin-Teller line.}
  \label{fig:scaling_dim}
\end{figure}

The scaling dimensions are plotted in Fig.~\ref{fig:scaling_dim}.
Below the multicritical point, i.e., along the line of continuous phase transitions between the columnar ordered and disordered fluid states, the scaling dimensions $x_\text{nem}$ and $x_t$ are seen to continuously change with $v$, while $x_\text{col}$ remains constant at a value consistent with the theoretical expectation of $x_\text{col} = 1/8$.  We have checked that the corresponding exponents $\eta_2$ and $\nu$ satisfy the Ashkin-Teller relation $\eta_2 = 1-1/2\nu$, i.e., $x_t/4 = x_\text{nem}$ within our errors.
On the other hand, we have $x_t/4 \neq x_\text{nem}$ for $v\geq 0.8$.
This discontinuous change in the critical index ratio strongly suggests the change from the single transition to the two separate transitions, i.e., the existence of the multi-critical point.
The expected $\eta=\eta_2=1/4$ at the multicritical point is actually realized at about $v=0.7$.
Parenthetically, we note that we find a decoupled Ising point at about $v=0.3$ in the $z=0.2$ plane, where we have $x_\text{nem}=0.25$ and $x_t=1.0$.

The scaling dimension of the energy operator can also be estimated from the FSS analysis of the Binder ratio as shown in Fig.~\ref{fig:fss}. Although the nonmonotonic behavior of the Binder ratio for columnar order and the relatively closely spaced successive transitions makes this difficult, we obtain almost the same results for $x_t$ from this analysis, as we do using the earlier analysis in terms of the order parameters.
We emphasize that nonmonotonicity has to do with the differing character of the columnar order parameter fluctuations in the nematic and the disordered fluid phases, and the related presence of a proximate multicritical point. Similar nonmonotonicity has been noted in the $J_1$-$J_2$ Ising model earlier~\cite{Jin2012}. In contrast to other examples of such behavior in frustrated Ising models, which is associated with a proximate weakly first order transition, we do not find evidence of any first-order transitions for the values of $v$ studied here.

Although it is difficult to completely exclude possibility of the weakly first-order phase transition around the multicritical point in our simulations, the weight of evidence suggests that the presence of a sizable $v$ term replaces the first order transition found in Refs.~\cite{Alet2005,Alet2006} by an intermediate nematic phase flanked by two second-order Ising phase boundaries. This is consistent with the fact that generalized four-state clock models, which serve as a discrete hard-spin analog of the coarse-grained description in terms of order parameter fields $\psi$ and $\phi$, are known for some parameter values to have such an intermediate phase flanked by two Ising lines that meet at a multicritical point with enhanced four-state Potts symmetry at the end of a line of Ashkin-Teller transitions~\cite{AlcarazKoberle1980}.

The multicritical point at $v=v_c(z)$ is expected to have the four-state Potts universality.
It is difficult to determine accurate location of the multicritical point because of the finite-size or finite bond-dimension effect.
To make matters worse, one also expects that a logarithmic correction appears at the four-state Potts model.
Our data, however, support the $S_4$ symmetry at the multicritical point.
The order parameters satisfy $m_\text{nem} > m_\text{col}$ below $v_c$, while $m_\text{nem} > m_\text{col}$ for $v>v_c$ (Fig.~\ref{fig:order_param}).
Thus we expect that $m_\text{col} \simeq m_\text{nem}$ at $v=v_c$, which indicates the emergent $S_4$ symmetry.

\begin{figure}
  \centering
  \includegraphics{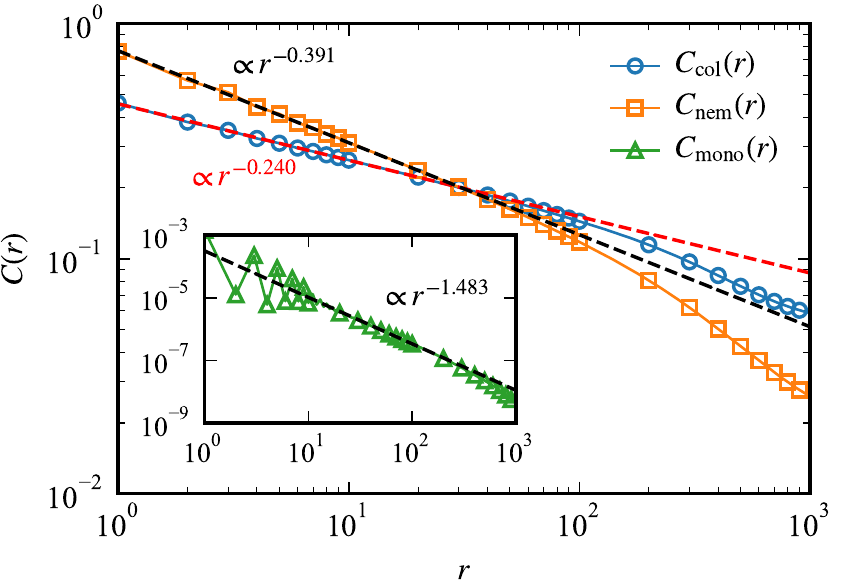}
  
  \caption{The correlation functions at $T=0.77138$ and $(u, v, z)=(0.4, 0.6, 0.2)$ calculated by the TN method with $\chi=128$. The monomer-monomer truncated correlation function is shown in the inset. The dashed lines are obtained by the linear fitting.}
  \label{fig:cf}
\end{figure}

By definition, the scaling dimension also appears in the corresponding critical two-point correlation function as $C_{a}(r)\propto r^{-2x_a}$.
We consider the correlation function along an axis because the uMPS can easily calculate it.
The correlation function between the local quantities is defined as
\begin{equation}
  C_a(r) \equiv \frac{1}{N} \sum_{\bm{\rho}}
  \left<
    \Psi_a(\bm{\rho}) \Psi_a^*(\bm{\rho} + r\bm{e}_{\alpha})
  \right>.
\end{equation}
In our TN simulation, $\bm{\rho}$ runs over sites in the $2\times 2$ unit cell, which corresponds to a value of $N=4$.
For correlation between monomers, the truncated correlation function,
\begin{multline}
  C_\text{mono}(r) \equiv
  \frac{1}{N} \sum_{\bm{\rho}} \bigl\{
  \left<
    n_\text{m}(\bm{\rho}) n_\text{m}(\bm{\rho} + r\bm{e}_{\alpha})
  \right> \\
  -
  \left<
    n_\text{m}(\bm{\rho})
  \right>
  \left<
      n_\text{m}(\bm{\rho} + r\bm{e}_{\alpha})
    \right> \bigr\},
\end{multline}
is expected to scale as $r^{-2x_t}$.
As shown in Fig.~\ref{fig:cf}, the scaling dimensions extracted by the linear fitting of the correlation functions agree with the results obtained by the FSS analysis.

The central charge is another important universal property of a critical point.
According to the conformal field theory, the correlation length $\xi$ and the entanglement entropy $S_\text{EE}$ are related by the Calabrese-Cardy formula at criticality:
\begin{equation}
  S_\text{EE} = \frac{c}{6} \ln \xi + \text{const.},
  \label{eq:Calabrese-Cardy}
\end{equation}
where $c$ denotes the central charge.

\begin{figure}
  \centering
  \includegraphics{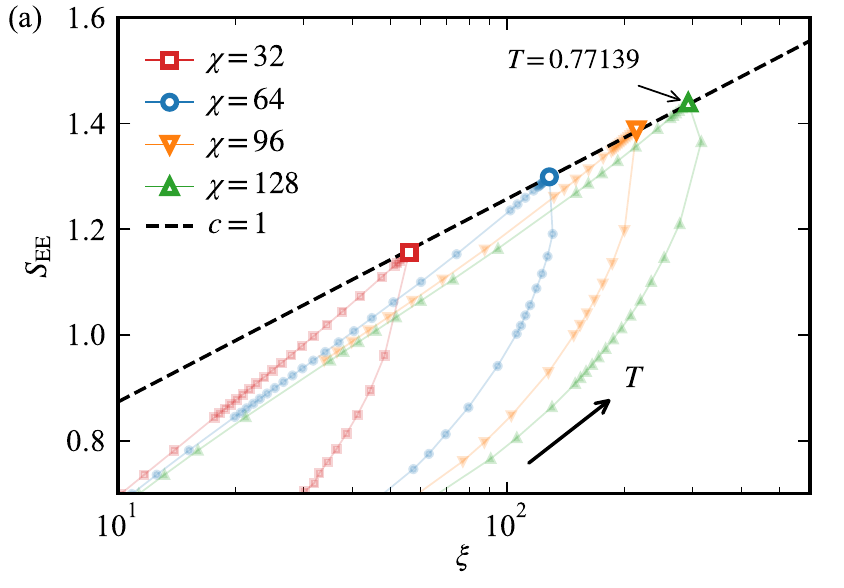}
  \includegraphics{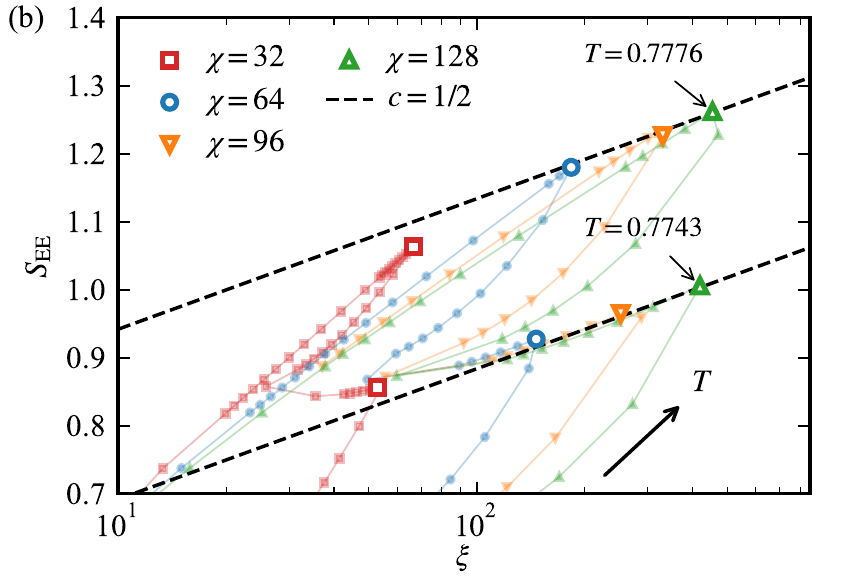}
  \caption{The entanglement entropy $S_\text{EE}$ vs the transverse correlation length $\xi_\perp$ at (a) $(u, v, z)=(0.4, 0.6, 0.2)$ and (b) $(0.2, 0.8, 0.2)$. The peaks of $S_\text{EE}$ are shown by a large symbol. The dashed lines are guides to the eye and its slope is equal to $c/6$ corresponding to the Calabrese-Cardy formula
  \eqref{eq:Calabrese-Cardy}.}
  \label{fig:S_vs_xi}
\end{figure}

One of the advantages of the TN method is that these two quantities can be calculated naturally.
Figure \ref{fig:S_vs_xi} clearly shows that the values obtained from the TN method are consistent with the Cardy-Calabrese formula \eqref{eq:Calabrese-Cardy}.
Based on the theoretical framework outlined in the previous section, the continuous phase transition between the columnar ordered and disordered fluid phases is expected to have a central charge of $c=1$.
On the other hand, the continuous Ising phase boundaries of the nematic phase are expected to have a central charge of $c=1/2$. From the results displayed in Fig. \ref{fig:S_vs_xi}, we see that our results do indeed conform to both these expectations.

\begin{figure}
  \centering
  \includegraphics{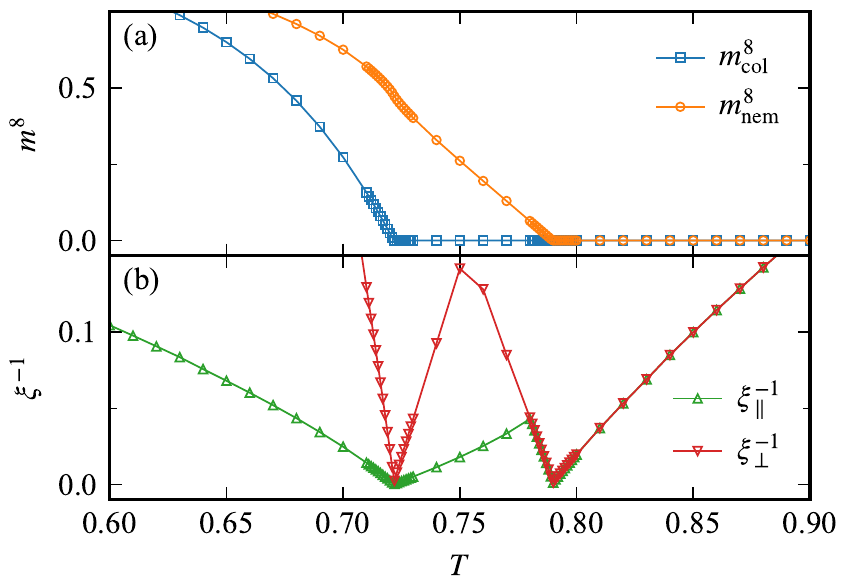}
  \caption{(a) The eighth power of the order parameters and (b) the inverse of the correlation length at $(u, v, z)=(0.1, 0.9, 0.2)$. The linearity of them around the transition points agrees with the Ising universality class.}
  \label{fig:m8}
\end{figure}

Above the multicritical point ($v> v_c$), there are two phase transitions, the columnar-nematic and nematic-disorder ones.
Although the FSS result of $x_t$ deviates from the expected value (Fig.~\ref{fig:scaling_dim}), we believe that it is due to the correction to scaling and effects from another critical point.
The TN result at $v=0.9$ shows that the eighth power of the order parameters and the inverse of the correlation length are linear to the temperature near the criticality (Fig.~\ref{fig:m8}).
This fact strongly indicates that these critical exponents satisfy $\beta=1/8$ and $\nu=1$, which is consistent with the Ising universality.

Finally, we comment on the approach of the multicritical point to $z=0$ as $v/u$ is increased. This happens very rapidly, and simultaneously, the phase boundary between the columnar and nematic phases, shown by orange squares in Fig.~\ref{fig:phase_zT}, approaches  the vertical temperature axis as $v/u$ increases. This is clear from monitoring the relative values of $m_\text{col}$ and $m_\text{nem}$ as in our earlier discussion about the symmetry of the multicritical point.
Eventually, in the limit $v/u\rightarrow\infty$, the columnar phase vanishes and only the phase boundary between the nematic and disordered phases remains.

\section{Discussions and Summary}\label{sec:summary}

In this paper, we have studied the classical grand-canonical monomer-dimer model on the square lattice with two types of attractive interactions.
We have determined the phase diagrams and analyzed the nature of the phase transitions using MC and TN methods.
The phase transition between disordered and columnar ordered phases shows the same feature as the Ashkin-Teller transition, where the critical exponents change continuously.
On the other hand, both nematic-disordered and columnar-nematic phase transitions belong to the Ising universality class.
Our numerical results show that the multicritical point, where three phases meet, has a positive fugacity when $v/u< \infty$.

Our conclusions and the theoretical framework within which they are situated has already been discussed at length. Here, we confine ourselves to highlighting one aspect of the phase diagram that appears to be worth further study. This has to do with the rapidity with which the multicritical point (at which the two Ising transitions meet the Ashkin-Teller line) moves towards $z=0$ as we increase $v/u$. The proximity to the full-packing limit raises the possibility that aspects of this could be understood by expanding about the full-packing limit in some way. It would be interesting to explore this in future work.
A related question has to do with the extent of the nematic phase itself in the $T$--$z$ phase diagram at large $v/u$. As noted in Ref.~\cite{Papanikolaou2014}, one expects that the low-temperature phase at full-packing will be columnar ordered for any finite value of $v/u$, no matter how large. However, the extent of this phase in $z$ decreases very rapidly with increasing $v/u$, until, at asymptotically large values of $v/u$, the columnar state only exists at full packing. Again, it would be interesting if some small $z$ expansion method could yield a more quantitative characterization of this phenomenon, which is very challenging to study by numerical methods.

\begin{acknowledgements}
  The work of S.M., H.-Y.L., and N.K. was supported by JSPS KAKENHI Grants No.~JP19H01809 and No.~JP20K03780.
  K.D. was supported at the Tata Institute of Fundamental Research by DAE, India and in part by a J. C. Bose Fellowship (JCB/2020/000047) of SERB, DST India, and by the Infosys-Chandrasekharan Random Geometry Center (TIFR).
  The inception of this work was made possible by a Visiting Professor appointment (K.D.) and associated research support at the Institute for Solid State Physics (ISSP), the University of Tokyo.
  The computational component of this work has been done using the facilities of the Supercomputer Center, ISSP.
\end{acknowledgements}

\bibliography{uv-dimer}

\end{document}